\documentstyle[sprocl,epsf]{article}
\def\Journal#1#2#3#4{{#1} {\bf #2}, #3 (#4)}

\def\NPB{{\em Nucl. Phys.} B}
\def\PLB{{\em Phys. Lett.}  B}
\def\PRL{\em Phys. Rev. Lett.}
\def\PRD{{\em Phys. Rev.} D}

\newcommand {\grav}    {\rm{\tilde G}}

\newcommand {\Zzero}   {{\rm Z}^0}
\newcommand {\MZ}      {\rm{M_Z}}
\newcommand {\ee}         {{e^+e^-}}

\newcommand {\eeeeg}      {e^+e^-\rightarrow e^+e^-(\gamma)}

\newcommand {\eenng}      {e^+e^-\rightarrow \nu\bar{\nu}\gamma}
\newcommand {\eeGrGr}      {e^+e^-\rightarrow \tilde{G}\tilde{G}}
\newcommand {\eeGrGrg}      {e^+e^-\rightarrow \tilde{G}\tilde{G}\gamma}

\def\gt{\raisebox{0.2ex}{$>$}}

\begin{document}
\rightline{DFPD 99/EP/15}



\title{
SENSITIVITY TO THE GRAVITINO MASS  FROM SINGLE-PHOTON SPECTRUM
AT TESLA LINEAR COLLIDER \footnote{Contribution to the International Workshop on Linear Colliders, Sitges, Spain, April-May  1999.}.}

\author{  P. CHECCHIA}

\address{ I.N.F.N. sezione di Padova, via Marzolo 8 35131 Padova, Italy}

\maketitle\abstracts{
The spectrum of single-photon events detected in the forward 
($|cos\theta_{\gamma}|<0.98$)
and in the barrel  region
of a TESLA linear collider detector
was studied in order to investigate the process $\eeGrGrg$.
}







%

\section{ Introduction}
A superlight gravitino $\grav$ (several
orders of magnitude lighter than the eV) is predicted by supersymmetric 
models 
\cite{ref:GMSB}.
In the last years, the possibility of detecting a light gravitino in 
accelerator experiments was studied in detail~\cite{ref:bfz}, 
the cross-section for the process $\eeGrGrg$ was computed 
and experimental values were obtained by LEP 
experiments \cite{lep}.
According to~\cite{ref:bfz} this cross-section can be very large 
if the gravitino is sufficently light, independently of the masses of the 
other supersymmetric particles.
Furthermore, in case all other supersymmetric particles are too heavy to be
produced, the $\eeGrGrg$ could be the only signal of 
Supersymmetry at an $e^+e^-$ collider.

The cross-section for the invisible reaction $\eeGrGr$ is~\cite{ref:bfz}: 
\begin{equation}
\sigma_0 \equiv \sigma(\eeGrGr )=\frac{s^3}{160 \pi \left| F \right|^4},
\label{sigma0}
\end{equation}
where $F$, defining the supersymmetry-breaking scale $\Lambda_S=|F|^{1/2}$,
is related to the gravitino mass by
$
 |F|=\sqrt{3} m_{3/2}M_P,
 ~M_P\equiv (8 \pi G_N)^{-1/2} \simeq 2.4\times 10^{18}$ GeV.
The visible radiative production ($\eeGrGrg$)
is given  in terms of
 the double 
differential cross-section $d^2\sigma/(dx_{\gamma},dcos\theta_{\gamma})$ 
where $x_{\gamma}$ and $\theta_{\gamma}$
are the fraction of the beam energy
carried by the photon and the photon scattering angle 
with respect to the electron direction, respectively.
For the dominant soft and collinear part of the photon spectrum 
$(x_{\gamma}\ll~ 1,~ sin\theta_{\gamma}\ll~ 1)$ an estimate of the 
signal cross-section can be obtained by applying the 
standard approximate photon radiation formula
to the lowest order $\eeGrGr$ cross-section $\sigma_0$
at $s^{\prime}=(1-x_{\gamma})s$:

\begin{equation}
\frac{d^2\sigma}{dx_{\gamma},dcos\theta_{\gamma}} \simeq 
\sigma_0 [s^{\prime}]  \cdot \frac{\alpha}{\pi} 
\frac{1+(1-x_{\gamma})^2}{x_{\gamma}sin^2 \theta_{\gamma}}.
\label{sigmadiff}
\end{equation}

Since the total cross-section can be parametrised as:
\begin{equation}
\sigma =\frac{\alpha s^3}{320 \pi^2 |F|^4}\cdot I,
\label{sigmaint}
\end{equation}
where 
$
I\simeq\int_{x_{\gamma}^{min}}^{x_{\gamma}^{max}} dx_{\gamma} 
  \int_{\left| cos\theta_{\gamma}^{min}\right|}^{\left| cos\theta_{\gamma}^{max}\right|}
  4(1-x_{\gamma})^3  \cdot 
\frac{1+(1-x_{\gamma})^2}{x_{\gamma}sin^2 \theta_{\gamma}} dcos\theta_{\gamma},
$
it is evident that, in order to achieve the largest sensitivity for the signal,
the measured spectrum should cover as much as possible the region of
low energy and low polar angle  photons.
On the other hand, this region is also populated by 
the background from radiative $\eeeeg$ events where both 
leptons escape undetected along the beam pipe.
In order to suppress such a background 
a cut on the photon transverse energy is necessary.
In the TESLA environment, this cut should take into account  the
Luminometer capability to veto scattered electrons/positrons in presence
of the machine background.  

An irreducible physical background is due to  single-photons from the 
process $\eenng$ which have a polar angle distribution 
similar to the signal but in the energy spectrum 
they present the characteristic $\Zzero$ return peak
at $x_{\gamma}=1- \MZ^2/s$.
As a consequence, an evidence of $\eeGrGrg$
will show-up as an excess of events over the $\eenng$ background 
in the region far from the radiative  $\Zzero$ return.
%
Since the signal cross-section~(\ref{sigmaint})
grows as the sixth power of the centre-of-mass energy, 
the signal sensitivity is dominated by the 
the highest energy data.

In case no signal is detected, 
a limit $\sigma_l$ on the  cross-section~(\ref{sigmaint})
would correspond to a
lower limit on the gravitino mass~\cite{ref:bfz}:
$
 m_{3/2}~\gt~3.8 \cdot 10^{-6} eV  \left[ \frac{\sqrt{s}(GeV)}{200} \right]^{3/2} 
                                   \left[ \frac{I}{\sigma_l} \right]^{1/4}.
$

This note  evaluates the sensitivity to this process for two possible
high luminosity runs ( 500 fb$^{-1}$) at the TESLA Linear Collider 
at $\sqrt{s}=500$ and 800 GeV.

\section{Apparatus and event selection}
A  basic description of a possible detector for TESLA 
can be found in \cite{ref:cdr}. 

The present sensitivity evaluation 
is based on the measurement of the electromagnetic energy
clusters 
in the Forward and in the Barrel Electromagnetic Calorimeter,
and in the Luminometer, 
as well as on the capability of vetoing 
the charged particles using the tracking devices.
An event is selected as single-photon candidate 
if it satisfies the following criteria:

- one electromagnetic energy cluster with $E_\gamma>10$ GeV,  
      $x_{\gamma}<0.7$
      and  $\theta_\gamma>11^{\circ}$;

- no energy released in the Luminometer above the
 machine background level;

- no charged tracks;

- the transverse momentum detected in the calorimeter 
incompatible with the presence
of two beams both missing the Luminometer:
  $p_t> 0.025                  \cdot(\sqrt{s}- E_{\gamma})$ GeV.

The $p_t$ cut together with the Luminometer veto is adopted in order to remove 
the background from radiative $\eeeeg$ events with both 
leptons undetected in the beam pipe.
 As it can be seen in table~\ref{tab:acc},
the relevant detector for the acceptance is the Luminometer whose minimum 
angle defines the minimum transverse momentum required for the photon detected
in the Forward or Barrel calorimeter.
The minimum photon energy the calorimeters are able to detect
is therefore not relevant for the signal sensitivity and then the 
analysis can be performed with very high efficiency 
with any kind of Electromagnetic
Calorimeter.
It is  assumed that the Hadron Calorimeter and Scintillator informations 
can be used to get rid of cosmic ray background.

\begin{table}[bth]
\caption[]{ The  acceptance $I$ for the radiative photon
spectrum as function of 
several cut parameters for $\sqrt{s}=800$ GeV. 
The dominant influence of the $p_t$ cut is evident.
}
\vspace*{0.2cm}
\begin{center}
\begin{tabular}{|c|c|c|c|}
\hline
$I $   & $E_{\gamma}^{min}$ & $\theta_{\gamma}^{min}$ &$p_t/(\sqrt{s}- E_{\gamma})$ \\
\hline
  8.23           & 10 & 11 & 0.038 \\
\hline
  10.63           & 10 & 11 & 0.030 \\
\hline
  12.70           & 10 & 11 & 0.025 \\
  12.70           &  6 & 11 & 0.025 \\
  12.68           & 20 & 11 & 0.025 \\
  12.44           & 10 & 13 & 0.025 \\
  12.13           & 10 & 15 & 0.025 \\
\hline
  15.46           & 10 & 11 & 0.020 \\
\hline
\end{tabular}
\label{tab:acc}
\end{center}
\end{table}


\section{Expected results}

The detection efficiency expected with the  selection criteria defined above 
depends on the detector details but it can be conservatively assumed to be 
similar to that of the LEP experiments (the ALEPH efficiency
$\epsilon_{\gamma}=77 \%$ is used). Since the results depends
on the efficiency with a power 1/4, this assumption does not influence
significantly the final sensitivity.     
 
The main background $\eenng$
cross-section  has been computed  with KORALZ~\cite{ref:koralz}
and NUNUGPV~\cite{ref:nunugpv} programs 
at 500 and 800 GeV \footnote{ The 
precision of both programs at the Linear Collider energies 
is of the order of 
$ 20\% \div 30 \%$. The author wish to thank A. de Min for his help in the
Monte-Carlo data production.}
 and it is in the range $ 1.1\div 1.5 $ pb
inside the acceptance region.
The corresponding expected number of events 
with 500 $fb^{-1}$ is then 550 $\div$ 750$\times 10^3$ events 
and hence the relative statistical error would be in the range
1.1$\div$1.3$~\times$10$^{-3}$.

\begin{figure}[hbt]
\begin{center}\mbox{\epsfxsize 6cm\epsfbox{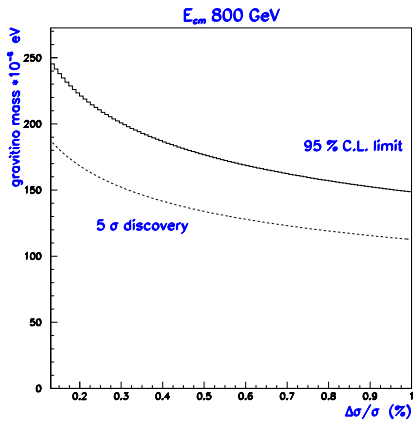}} \end{center}
\caption{ Sensitivity to $\grav$ mass as function of the relative error 
on the $\eenng$ background cross-section at 800 GeV centre-of-mass Energy.} 
\label{sens}
\end{figure}

Given the high background rate, the systematic uncertitude 
on its measurement  determines the sensitivity for the
signal detection. Assuming the experimental systematics being
dominated by the absolute Luminosity determination 
(a conservative value for $\Delta {\cal L}/{\cal L}$=1$\%$  
was given in~\cite{ref:cdr})
the sensitivity as function of the relative error 
$\Delta \sigma/\sigma$ on the  $e^+e^- \to \nu \nu \gamma$ cross-section  
measurement is shown in Fig.~\ref{sens} for a $95\%$ Confidence Level limit
and for a 5 standard deviations discovery.
For a total error of about $0.5 \%$
the limit at 800  (500) GeV would be
$m_{3/2}>1.8~(0.8) \times 10^{-4}$ eV at 95 $\%$ C.L..
This error value is compatible with the expectation for the 
future theoretical precision on the cross-section computation \cite{ref:nicro}.
The analysis can be improved by comparing the measured energy spectrum with 
the background and signal+background expected ones by means of the 
Likelihood Ratio technique \cite{ref:read}.   

\section{Conclusions}
The channel $ e^+e^-\rightarrow \gamma X_{invisible}$ 
is a  very important tool for investigating new Physics.
Excess of events 
in the low energy part of the photon spectrum for the highest $\ee$ 
energy 
could be due to  superlight $\grav$ production or, 
as recently claimed~\cite{ref:graviton},
to Extra Dimensions in Quantum Gravity. 
Unfortunately the presence of 
a very high backgrond from the $\eenng$
channel requires  to measure the single photon cross-section 
with a very high accuracy and to compute the expected cross-section 
with high precision in order not to spoil the sensitivity.

A preliminar evaluation of the sensitivity to the gravitino mass 
for a machine delivering
about 500 fb$^{-1}$ at $\sqrt{s}=800$ GeV
is $m_{3/2} \sim 1.5\div 2.3 \times 10^{-4} $ eV
corresponding to
 $\sqrt{\left|F \right|}\sim 0.8 \div 1 $ TeV.

The same experimental analysis can be applied  to the 
search for Quantum Gravity Extra Dimensions.


\section*{References}

\end{document}